\documentclass[twocolumn,prl,superscriptaddress,showpacs,floatfix]{revtex4}
\usepackage{graphicx}
\usepackage{epsfig}
\usepackage{rotating}
\usepackage{amsmath}
\usepackage{amsfonts}
\usepackage{amssymb}
\usepackage{enumerate}
\usepackage{longtable}
\usepackage{dcolumn}% Align table columns on decimal point
\usepackage{bm}
\DeclareMathOperator{\Tr}{Tr}

\begin{document}

\title{Confinement, reduced entanglement, and spin-glass order in a random quantum spin-ice model
}

\author{Anirudha Menon}
\affiliation{Department of Physics, University of California Davis, CA 95616, USA}

\author{Tom Pardini}
\affiliation{Lawrence Livermore National Laboratory, 7000 East Ave. Livermore, CA 94550}

%\author{Stefan P. Hau-Riege}
%\affiliation{Lawrence Livermore National Laboratory, 7000 East Ave. Livermore, CA 94550}

\author{Rajiv  R. P. Singh}
\affiliation{Department of Physics, University of California Davis, CA 95616, USA}

\date{\rm\today}

\begin{abstract}
	We study an effective spin model derived perturbatively from random transverse-field Ising model on the
	pyrochlore lattice. The model consists of spin-configurations on the pyrochlore lattice, restricted 
	to the spin-ice subspace, with spins interacting with random Ising exchange couplings as well as
	ring exchanges	along the hexagons of the lattice. This model is studied by exact diagonalization 
	upto N=64 site systems. We calculate
	spin-glass correlation functions and local entanglement entropy $S_T$ between spins in a single tetrahedron and
	the rest of the system. We find that the model undergoes two phase transitions. At weak randomness
	the model is in a quantum spin-ice phase where $S_T=\ln{6}$.
% implying that the subsystem is in an equal admixture of the six classical spin-ice configurations. 
	Increasing randomness first leads to a frozen phase, with long-range spin-glass order and $S_T=\ln{2}$
	corresponding to the Cat states associated with Ising order. Further increase in randomness
	leads to a random resonating-hexagon phase with a frozen backbone of spins and 
	a broad distribution of entanglement entropies.
	The implications of these studies for non-Kramers rare-earth pyrochlores are discussed.
\end{abstract}

\pacs{74.70.-b,75.10.Jm,75.40.Gb,75.30.Ds}

\maketitle

\section{Introduction}

The study of quantum spin-liquids has emerged as one of the key directions in condensed matter physics
\cite{Balents_nature,balents2,kagome,triangular,kitaev,kitaev2,khaliullin,banerjee, matsuda}.
It is driven in part by a desire to understand emergent quantum phases that defy conventional description
and in part by a hope that at some time in future one might be able to exploit topological excitations in next generation quantum devices. Indeed, 
on the theoretical end, there has been substantial progress. Exactly soluble models, such as the Kitaev
model \cite{kitaev}, dispel any doubts that emergent phenomena, with fractionalized quasi-particles and emergent
gauge theories can arise in a system of interacting spins. Computational approaches, most notably, area-law 
entanglement inspired density matrix renormalization group (DMRG) and tensor-network methods have greatly
improved our ability to simulate the behavior of such systems, while methods of quantum field-theory, including
parton constructions and topological field-theory have helped to catalog and systematize the range of relevant phenomena \cite{balents2}.

In contrast, progress has been relatively slow in convincingly demonstrating quantum spin-liquid behavior in
real materials. This is in spite of the fact that an abundance of candidate quantum spin-liquid
materials exist. These include the triangular and Kagome families of geometrically frustrated
two-dimensional materials \cite{kagome,triangular}, the highly anisotropic spin-orbit coupling dominated Kitaev materials \cite{khaliullin,banerjee,matsuda} and the 
spin-ice materials mostly magnetic rare-earth pyrochlores \cite{spin_ice_review,ramirez,ross,savary,jaubert}. Despite a few decades of concerted effort,
it has been difficult to unambiguously establish fractionalization and emergence in these systems. 
Part of the difficulty is simply that emergent fractionalized quasiparticles do not directly couple with
external probes, making their detection difficult. But, part of the difficulty is also associated with the 
existence of quenched impurities, which can strongly modify the behavior as well as obscure or mimic 
signatures of emergence \cite{kimchi,chernyshev2,perkins1,kawamura1,kawamura2,andrade,vbg,mila}.
How much impurities can a quantum spin-liquid tolerate and what kind of behavior results from quenched
impurities, remains an important question in the field.

In this work we study an effective model derived perturbatively from a random transverse-field Ising model (RTFIM) 
on the pyrochlore lattice \cite{savary1,benton}. RTFIM has been shown recently to be relevant for non-Kramers family of rare-earth 
pyrochlores \cite{savary1,benton,broholm}. Because
two-fold degeneracy for eigenstates of non-Kramers ions is not guaranteed by time reversal symmetry, effective 
two-level models for these systems have random transverse-fields coming from quenched disorder. We had recently
studied such a random transverse-field Ising models by exact diagonalization (ED) and numerical linked cluster expansion (NLC) methods \cite{PMHS}, finding
a phase diagram consisting of quantum spin-liquid (QSL), paramagnetic (PM), Ising (I) and Griffiths-McCoy (GM)
phases. The systems studied were too small to establish whether the Ising phases had long-range spin-glass order.

Here, we use perturbation theory to derive an effective model starting from RTFIM, where spins are restricted to the spin-ice
configurations \cite{savary1,benton}. In the spin-ice subspace there are two types of random interactions. There are 
Ising exchanges along the bonds and there are cyclic ring-exchanges along the hexagonal loops of the model.
Both these interactions are dependent on the configuration of random fields.
The effective model can be studied up to larger systems (N=64 spins) allowing us to clearly establish
the nature of the confined phases. We study a two-parameter family of models characterized
by a mean value $h$ and a width $w$ for transverse-fields. As the width is increased
at small $h$ values, one observes two phase transitions. First, into a confined phase with long range spin-glass
order and then into a partially deconfined phase, which we call a random resonating-hexagon
(RRH) phase. In the latter, the resonating hexagons are randomly placed and leave a fraction of 
spins forming a frozen spinglass backbone. We believe the latter phase is a variant of cluster glass phases
such as random singlet phase \cite{bhatt-lee,dsf,vbg,kimchi}
that arise naturally in random quantum spin models.
The implication of these findings to the rare earth pyrochlores will be discussed.

\section{Models and Methods}

We begin with the Hamiltonian for a transverse-field Ising model on the pyrochlore lattice \cite{moessner,tfim1,tfim2,gchen}:
\begin{equation}
{\cal H}=J\sum_{<i,j>} \sigma_i^z \sigma_j^z
- \sum_i h_i \sigma_i^x,
\end{equation}
where the sigmas denote Pauli spin matrices, and $J$ is the Ising exchange constant for nearest-neighbor interactions
on the pyrochlore lattice. 
The transverse fields $h_i$ are assumed to be independent random variables at each site. In this work we will take the
distribution of $h_i$ to be Gaussian with mean $h$ and standard deviation $w$, although details of the distribution
are not critical to the study.
Working in the limit of $J\to\infty$, restricts us to the Hilbert space of spin-ice states. In this space,
the first perturbations that lift the degeneracy of ice states are 
fourth order terms that give rise to an Ising coupling on
 a bond given by strength proportional to $h_i^2h_j^2$ \cite{benton}. Then, in $6$th order perturbation theory we get the ring exchange
 term on the hexagons that can lead to a resonating QSL state.

 In the spirit of effective models in a reduced subspace, like the $t-J$ model derived from the Hubbard model,
 we will now study the effective model in the spin-ice subspace:
\begin{eqnarray}
	{\cal H}=&{1\over 48} \sum_{<i,j>} h_i^2 h_j^2 \sigma_i^z \sigma_j^z \\ \nonumber
	&- {63\over 256} \sum_{u} K_u (\sigma_1^+\sigma_2^-\sigma_3^+\sigma_4^-\sigma_5^+\sigma_6^-\ +\ h.c.),
\end{eqnarray}
where $h_i$ are quenched random variables with mean $h$ and width $w$. The second sum is over all hexagons $u$ of
the lattice and $K_u=\prod_{i=1}^6 h_i$ where $h_i$ are the random fields at the six sites of the hexagon and
$\sigma_1$ through $\sigma_6$ are the spin operators on the hexagon in a cyclic order.
Note that the cyclic term is only operative when the spins alternate along the hexagon, otherwise it destroys the state.
From here onward we will study this effective model, which depends on two parameters $h$ and $w$, and will allow for all
values of $h$ and $w$.

For $w=0$, the Ising couplings do not cause any dispersion in the spin-ice subspace. Hence, the model reduces to the pure hexagonal ring exchange
model \cite{hermele,castro-neto} simulated by several groups before \cite{shannon,QED-numerics}. This model is known to have a quantum spin liquid ground state with emergent
quantum electrodynamics and a collective photon excitation. Our goal is to study different phases of the model
as a function of $w$ and $h$.

We compute the following quantities:

1. The many-body band-width of the system, per spin, defined as
\begin{equation}
	{B\over N} ={E_{max} - E_{min}\over N},
\end{equation}
where $E_{max}$ is the energy of the highest energy state and $E_{min}$ the energy of the lowest energy state
for an $N$-site cluster.

2. Entanglement entropy of a tetrahedron of spins and their distribution:
In a pure state, the von-Neumann entanglement entropy for subsystem $A$ and its complement $B$ is defined as
\begin{equation}
S_A=S_B = -\Tr \rho_A \ln{\rho_A},
\end{equation}
where $\rho_A$ is the reduced density matrix for subsystem $A$.
In this work, A is made up of the four spins belonging to any single tetrahedron.
%This leads to the definition of single-tetrahedron entanglement entropy, which we then averaged over different tetrahedron
%in the system and the disorder configurations.
%\begin{equation}
%S_T= -\Tr \rho_T \ln{\rho_T}.
%\end{equation}
In the uniform system, it is easy to see that the tetrahedron entanglement entropy
$S_T=\ln{6}$. Let us label our basis states as
$|\alpha,i>,$
where $\alpha$ stands for spins inside the tetrahedron and $i$ stands for the spins outside.
Then, a general state of the system can be expressed as
$$|\psi> = \sum_{\alpha,i} C_{\alpha,i} |\alpha,i>.$$
Reduced density matrix for the tetrahedron
$$\rho^T_{\alpha,\beta} = \sum_i C{\alpha,i}^*\ C_{\beta,i},$$
must be diagonal because for two state $|\alpha,i>$ and $|\beta, i>$ to both lie
in the spin-ice subspace, one must have $\alpha=\beta$. Furthermore, the uniform system
has sublattice symmetry and all $6$ states are related by a permutation of sublattices.
Hence, as long as there is a non-degenerate ground state, all $6$ diagonal matrix-elements of the
reduced density matrix must be equal implying $S_T=\ln{6}$. Our earlier exact diagonalization 
studies of $16$ and $32$ site systems \cite{PMHS} showed that, in the full random traneverse-field model, this entropy remains
very close to $\ln{6}$, despite virtual dressing of the spin-ice states due to perturbative
fluctuations. Only when one is near the confinement transition one sees deviations from this
value.

3. Ising correlation function and correlation sum
\begin{equation}
C_{ij}=[\langle \sigma_i^z \sigma_j^z \rangle ^2\  ],
\end{equation}
\begin{equation}
C_{zz}=[{1\over N}\sum_{i,j } \langle \sigma_i^z \sigma_j^z \rangle^2\  ],
\end{equation}
where the angular brackets refer to ground state averages and square brackets to an average
over the distribution of random fields.
Note that we have normalized the correlation sum such that
in a system with long-range (random) order the correlation sum $C_{zz}$ should scale as $N$.

4. Inverse participation ratio in the many-body space defined as
\begin{equation}
	IPR= {1\over \sum_i a_i^4},
\end{equation}
where $a_i$ are the coefficients of the ground state wavefunction in the Ising basis.
In the quantum spin-ice phase this quantity should be order $D$, where $D$ is the dimension of
the spin-ice space, whereas in an equal admixture Cat state of two Ising configurations, it should equal $2$.

\section{Numerical Results }

We study finite clusters with periodic boundary conditions. We have looked at clusters of size $16$, $32$, $40$, $48$
and $64$. 
%Each periodic cluster is characterized by three translational unit vectors, defining the repeating unit cell.
%In the notation of a cubic system and taking length of an elementary cube to be ..., the three translational vectors
%for the different clusters are ......
To carry out the calculations we first pick a configuration of random fields for each site from a
Gaussian random distribution with mean $h$ and uncertainty $w$. This allows us to determine the
Ising couplings for all the bonds as well as strength of the ring exchange for each hexagon.
The ground state is then obtained by the Lanczos method.
We typically include 
$100$  to $400$ different field configurations to average over random configurations.
Variations from different field configurations allow us
to determine the statistical error bars.

With only the ring exchanges as the off-diagonal terms, this model is known to partition into disjoint subspaces.
We first determine all the disjoint parts of the Hilbert space and
then obtain the ground state in each subspace. The overall ground state is obtained by comparing energies
in different subspaces.
In the QSL phase, the ground state typically lies in the largest connected
subspace but this is not necessarily true. When the ground state is in a subspace where a
state and its time reversed partner are disconnected, there must be
two degenerate ground states. Since, the original
transverse field model does not have a disconnected Hilbert space, in this case, we take as our ground state the symmetrized
linear superposition of the two.
For $N=64$ case, the total spin-ice subspace has dimensionality
$2,249,370$ whereas the largest disjoint sector has dimensionality $194,640$. 

In Fig.~1, the many-body band-width of the system per site is shown. It is independent of width $w$
in the QSL phase and depends on $h$ as $h^6$ due to the ring exchange term. 
In the perturbative regime, after the first phase transition the bandwidth goes as $w^2$ as known from previous 
studies  using NLC \cite{PMHS} for RTFIM. This follows from the fact
that within each tetrahedron the maximum difference in the energy between any two
spin-ice states goes as $w^2h^2$, which overwhelms
the $h^6$ term.
For large $w$ the bandwidth must scale as $w^6$.

In Fig.~2, we show the average entanglement entropy $S_T$ for a tetrahedron. It is averaged over
the tetrahedron in a cluster as well as over the distribution of random fields. Two phase
transitions are clearly seen in the plot. It equals
$S_T=\ln{6}$ in the QSL phase. In the spin-glass phase it approaches $\ln{2}$ implying
that the state is reduced to a cat state with only global $\ln{2}$ entropy and there is
negligible quantum fluctuation around that state. At very large
$w$ it approaches a value somewhat below $2\ln{2}$. 
Fig.~3 shows a distribution of entanglement entropies. In the transition regions as
well as in the random resonating-hexagon phase at large $w$ the entropy distribution becomes very broad.

In Fig.~4, the Ising correlation sum is shown. The correlation sum is extensive 
in both the confined phases as shown in Fig.~5. 
In the intermediate phase, which we identify as the Ising spin-glass, 
it approaches the maximal value, again showing that 
quantum fluctuations become negligible as the Ising couplings dominate over the ring exchanges.

In Fig.~6, the inverse participation ratio is plotted. Apart from the intermediate phase where it is
only slightly large than $2$, it is size dependent in the other two phases. In the QSL phase it must scale with the dimension
of the Hilbert space. In the random resonating-hexagon phase also it grows with the size of the system.

\begin{figure}
\begin{center}
 \includegraphics[width=7cm]{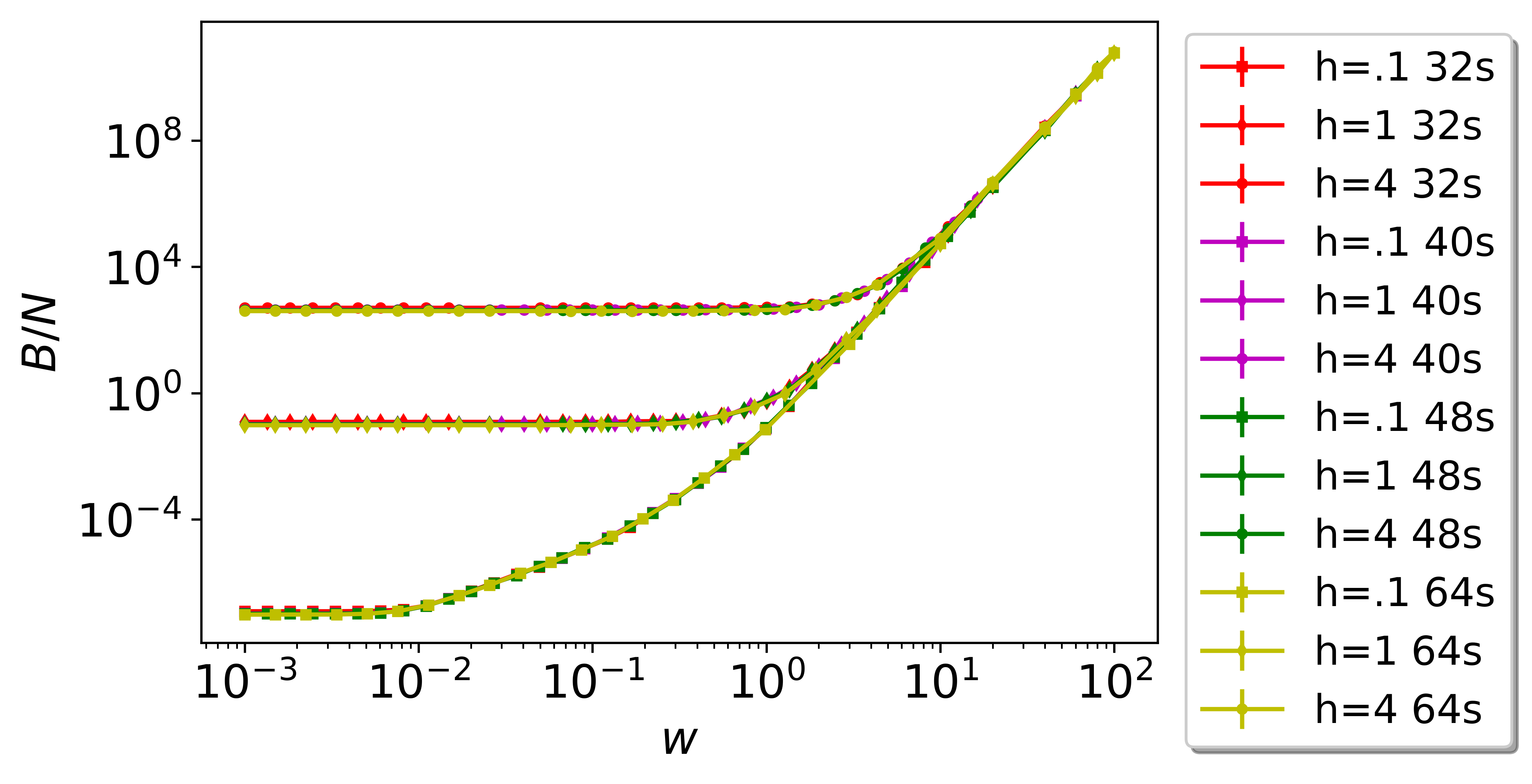}
\caption{\label{Bandwidth} 
The total Bandwidth of the spin-ice subspace per spin plotted on a log-log scale.
%	as a function of $w$ in the quantum spin-liquid (QSL) phase. It scales as $w^2$ in the Ising Spin-glass (ISG) phase
%	and it scales as $w^6$ in the random resonating-hexagon (RRH) phase. The intermediate phase is absent when $h$ is large.
	}
\end{center}
\end{figure}

\begin{figure}
\begin{center}
\includegraphics[width=6cm]{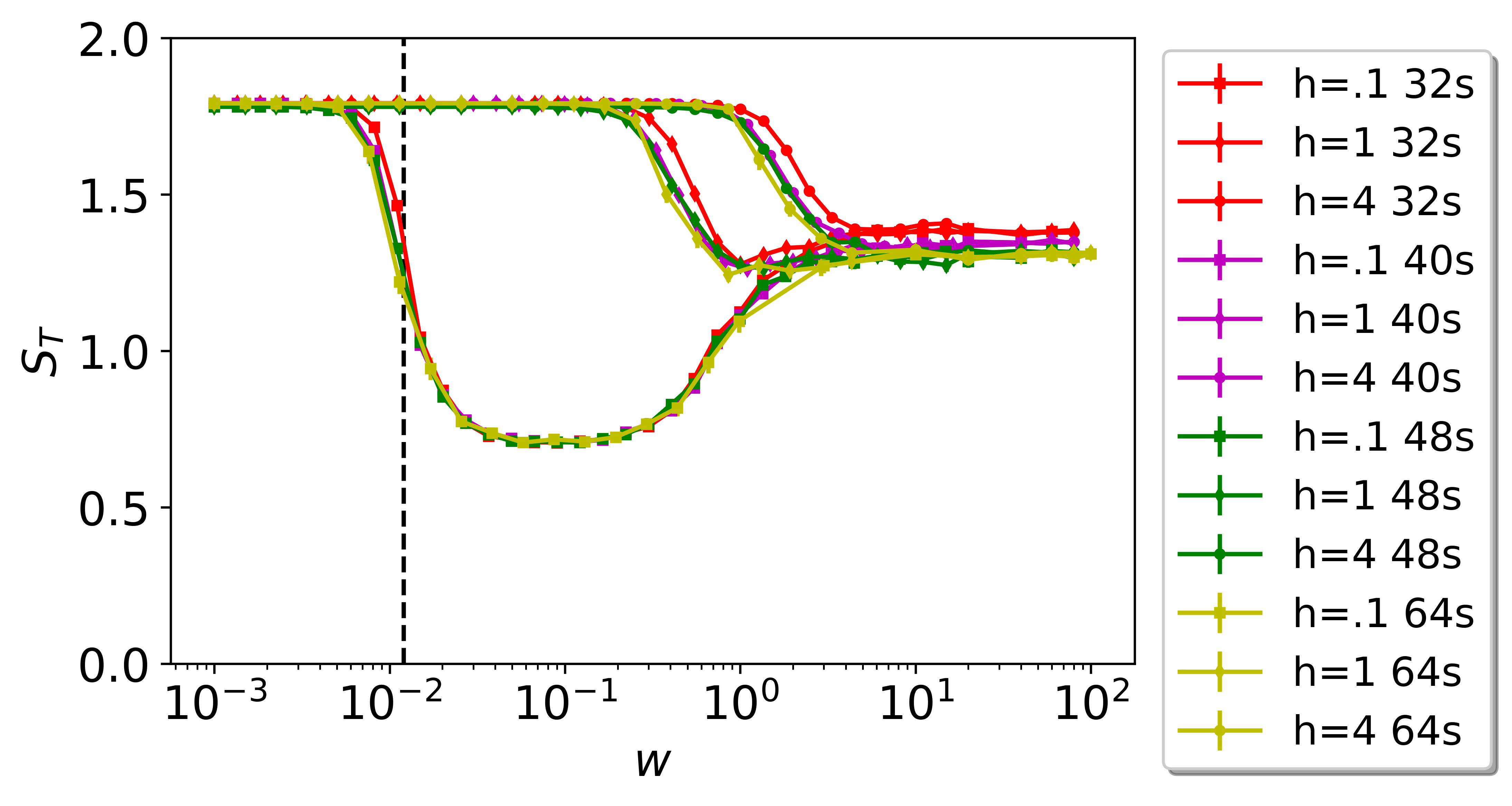}
\caption{\label{Entanglement} 
Entanglement entropy for a tetrahedron of spins, $S_T$, for different sizes and parameters.
	}
\end{center}
\end{figure}

\begin{figure}
\begin{center}
\includegraphics[width=6cm]{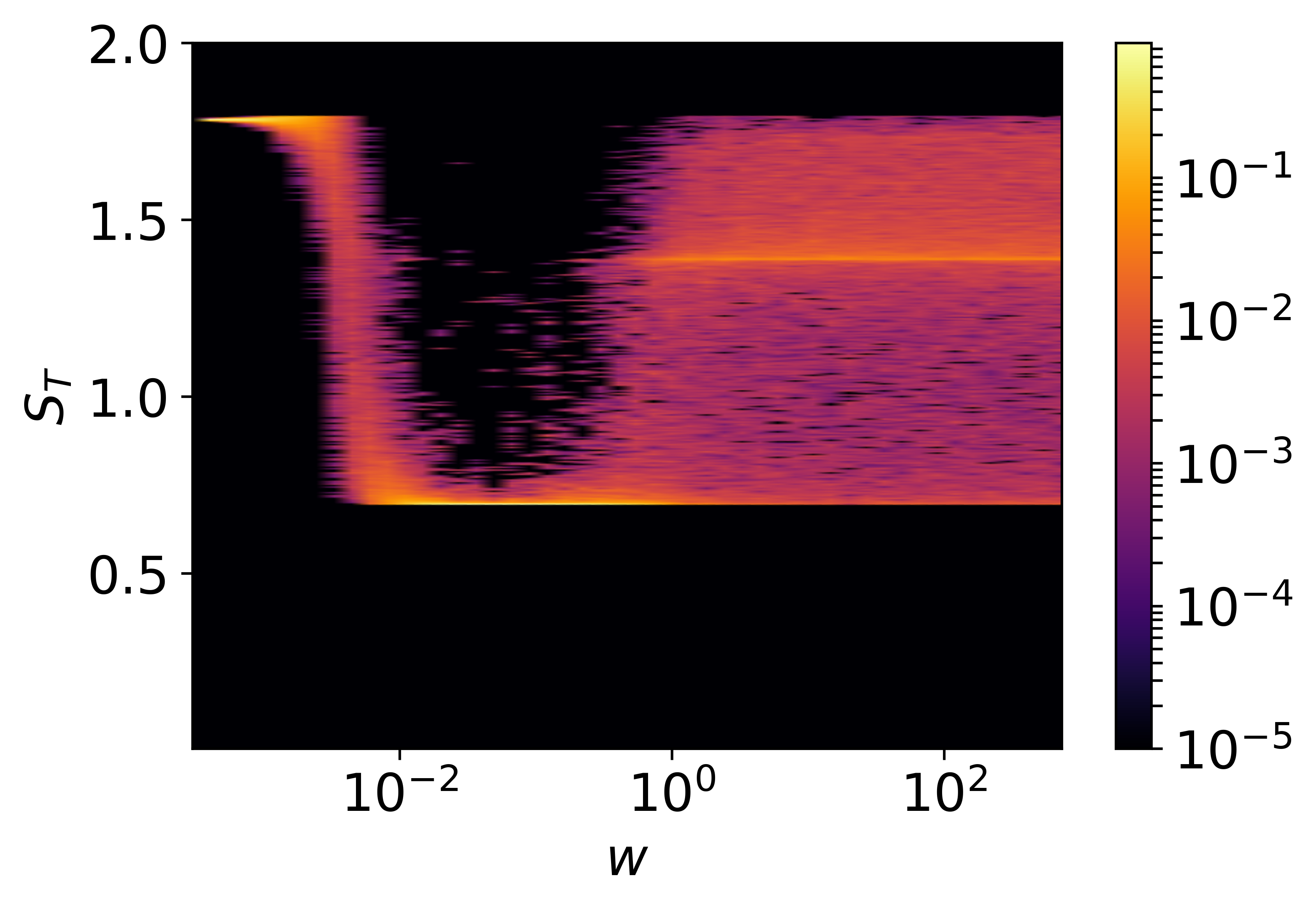}
\caption{\label{Distribution} 
	Distribution of entanglement entropies for different values of $w$ for $h=0.06$.
	}
\end{center}
\end{figure}

\begin{figure}
\begin{center}
\includegraphics[width=6cm]{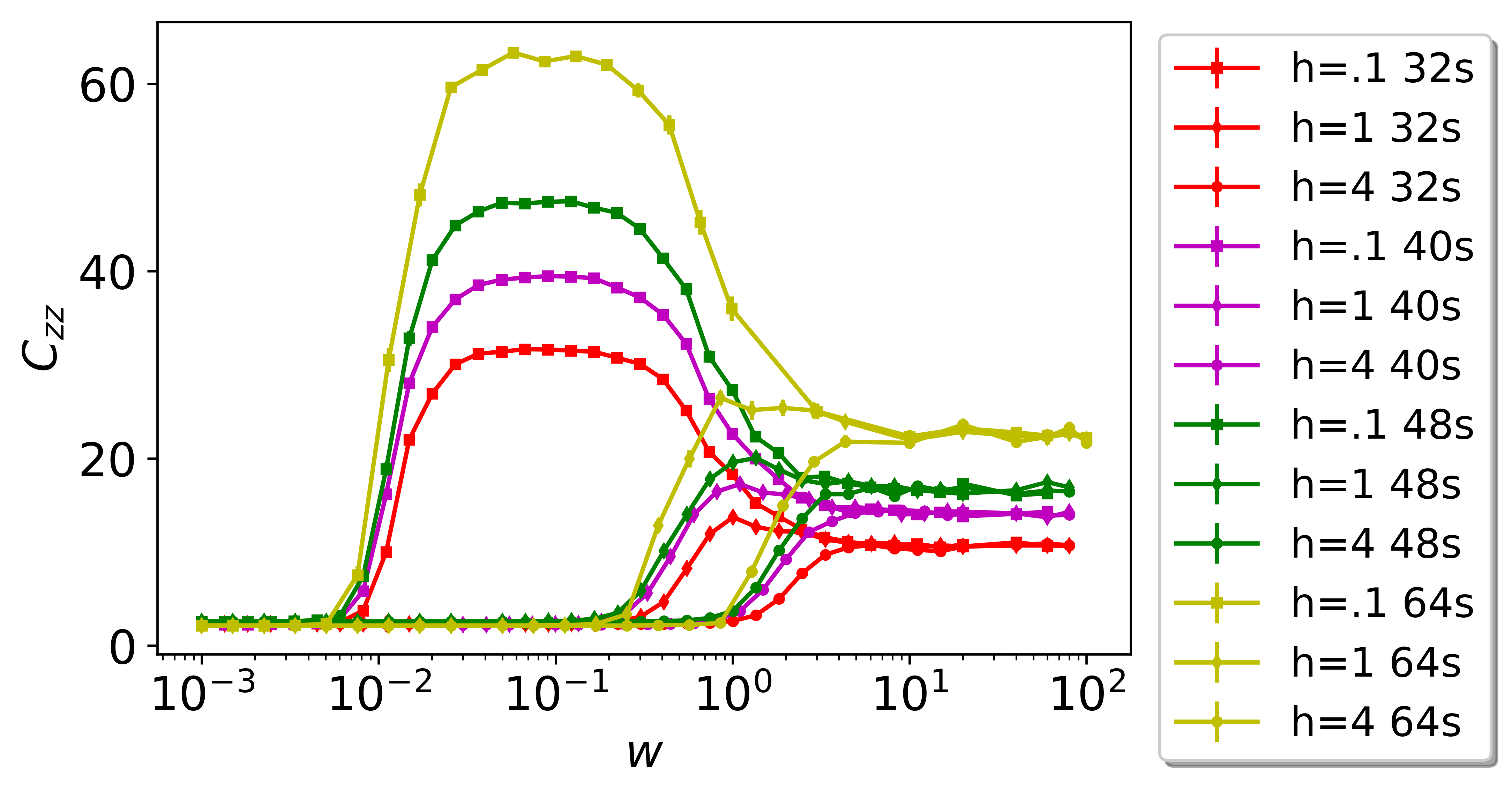}
\caption{\label{Ising} 
Ising correlation sum in different parameter regions.
	}
\end{center}
\end{figure}

\begin{figure}
\begin{center}
\includegraphics[width=6cm]{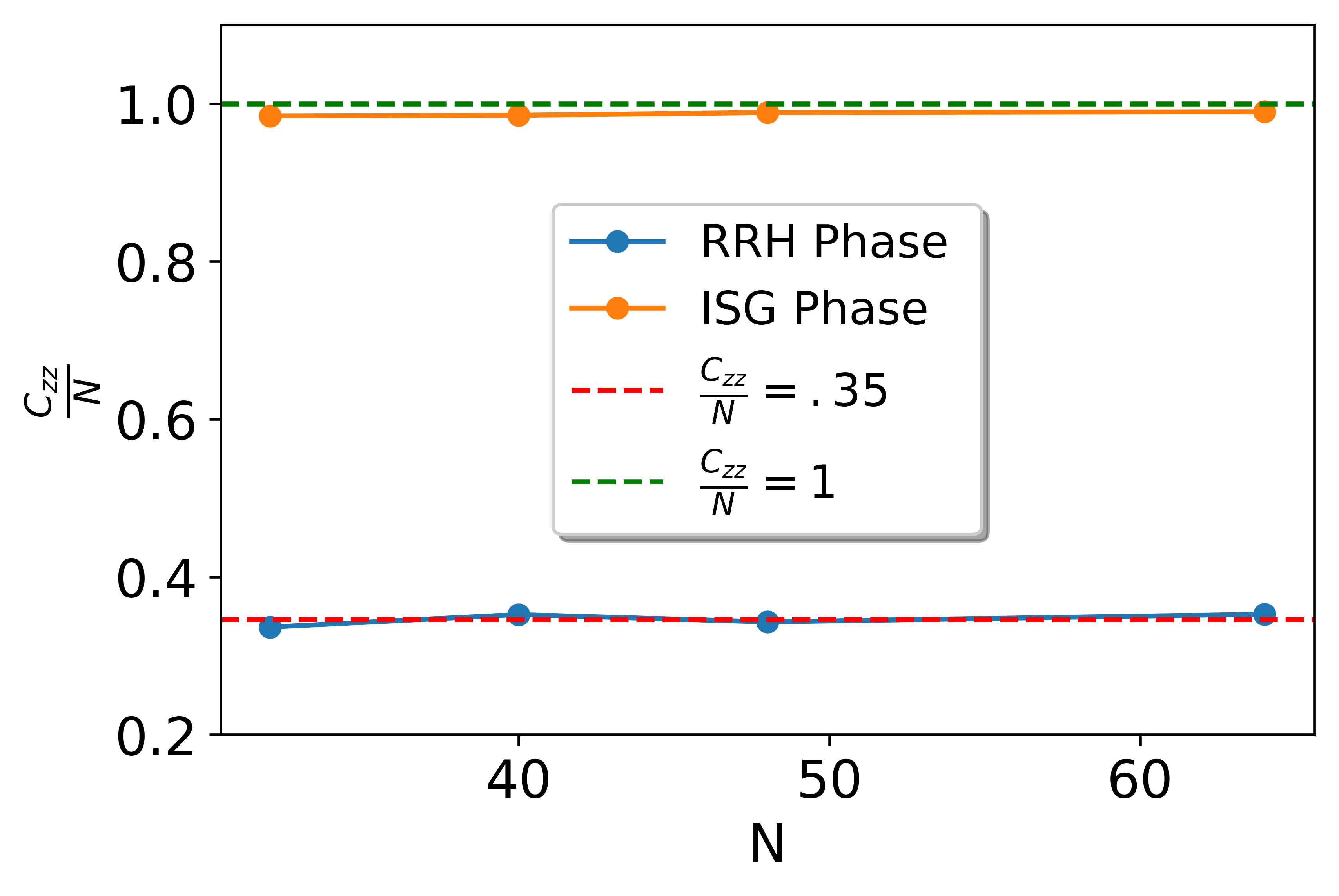}
\caption{\label{IsingvsN} 
Scaling of Ising correlation sum with $N$ the size of the system.
	}
\end{center}
\end{figure}

\begin{figure}
\begin{center}
\includegraphics[width=6cm]{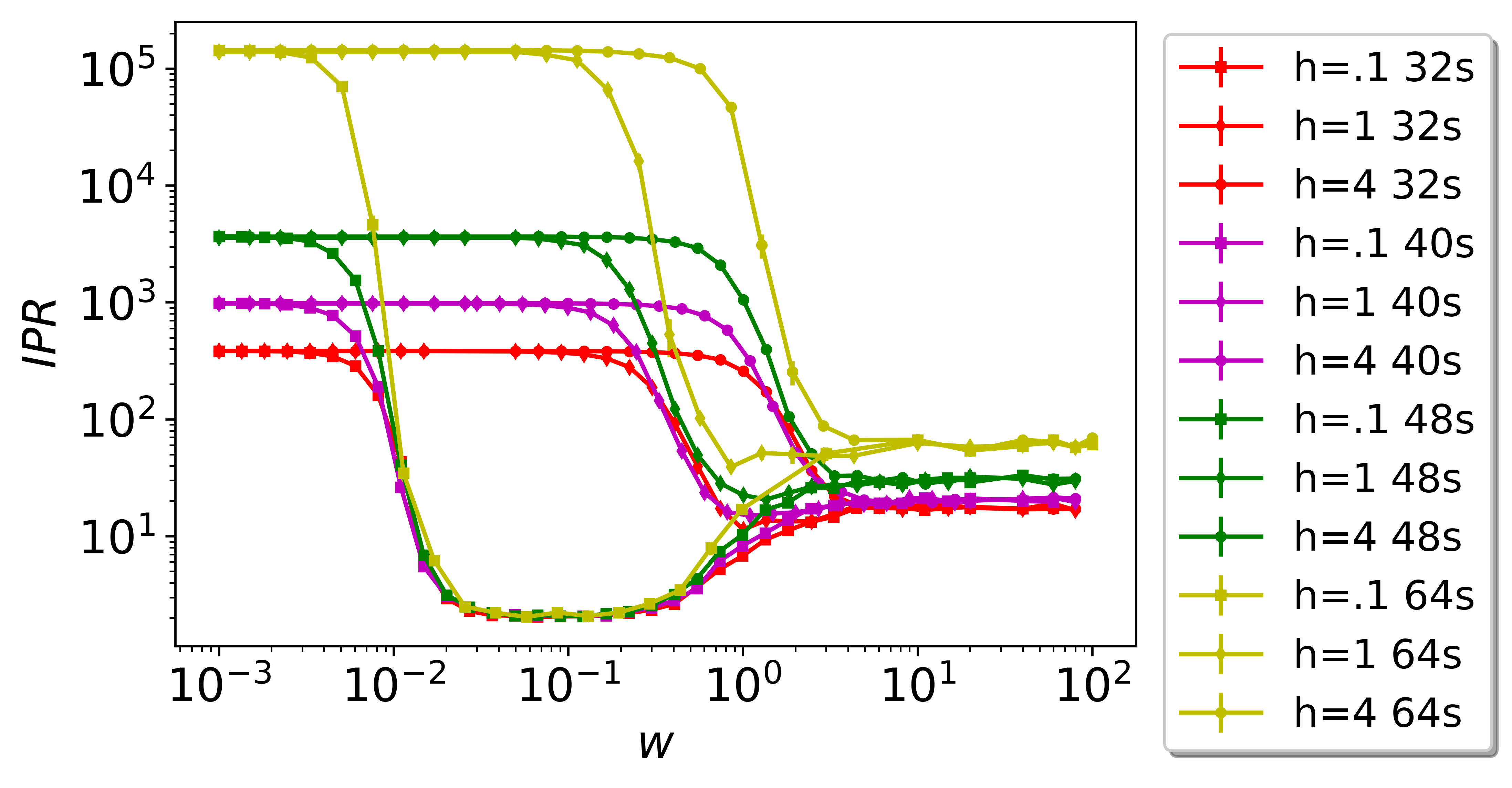}
\caption{\label{IPR} 
	Inverse Participation Ratio for different parameters.
	}
\end{center}
\end{figure}

\section{Discussions: Phase Diagram}
Our numerical results clearly establish three phases in the model separated by phase transitions.
As shown in the phase diagram in Fig.~7 these are the quantum spin-liquid (QSL), the Ising spin-glass (ISG)
and the random resonating-hexagon (RRH) phases. These phases are separated by sharp transition regions.
When there is no disorder, that is $w=0$, the model is in the QSL phase. This must be separated
from other phases by a thermodynamic phase transition as the other phases have long range Ising order
and hence a broken $Z_2$ symmetry. At small $h$ and $w$, the Ising Spinglass phase has negligible
quantum fluctuations. This is not surprising as the Ising couplings arise in lower order of
perturbation theory than the ring exchange term \cite{benton,PMHS}. Hence they totally dominate 
and as argued in our previous paper \cite{PMHS} all the ring terms are frozen out by disorder.
The transition is not a simple level crossing as clearly the transition region 
is broad. However,
we do not see evidence for crossing of physical properties for different sizes
as expected at a critical transition.
Instead there is a weak size dependence of the transition point. Yet, the transition region
is characterized by substantial fluctuations, and the entanglement entropy develops a broad distribution there.
Hence, we conclude that the transition is weakly first order. Savary and Balents \cite{savary1} suggested
a weakly first order confinement transition due to randomness even when no spin-glass order develops on the other side.
The development of an additional order is even more likely to make it first order.

The most interesting phase is the random resonating-hexagon (RRH) phase. This phase has a broad distribution 
of entanglement entropy with a peak near $2\ln{2}$. The many-body inverse participation ratio suggests that
it still grows exponentially with system size, although very slowly $1.04^N$. 
And, on top of that this phase has Ising correlation sum which scales as $N$ implying long-range order.
At large random-fields the Ising exchange terms become negligible and the model becomes a random ring-exchange
term. We believe the ground state of this model can be understood along the lines of real-space renormalization
group (RSRG) approaches to understanding random quantum spin problems \cite{bhatt-lee,dsf,vbg,kimchi}. As the distribution of couplings
becomes very broad (being a product of $6$ field terms), one can first pick out the strongest hexagon
and make it into a resonating cluster. The strong resonance of one hexagon makes these six spins ineligible for additional entanglement and renders neighboring spins frozen into certain configurations 
as otherwise they would interfere with this resonance \cite{PMHS}. This resonating cluster is like a singlet formation in a random-singlet phase,
except it involves six spins and is unrelated to any SU(2) symmetry.
Then one can pick out the next strongest resonating hexagon.
When this process is continued, it will lead to a random configuration of hexagons 
placed in the lattice and it can leave a backbone of spins that must be fixed in order to maintain
compatibility with the resonance in the hexagons and be in the spin-ice subspace. Our numerical results
suggest that the backbone is percolating and has a finite fraction of spins.

The participation ratio should be exponential in number of resonating clusters, which should be of order one
per ten to twenty spins. Furthermore, one would expect a large distribution of entanglement entropies in the
phase with a substantial weight around $2\ln{2}$, corresponding to the four spins in a tetrahedron dividing
into two groups of two and participating in two independent resonances.

Going back to the random transverse-field Ising model (RTFIM), such an RRH phase can only arise at intermediate values
of disorder $w$ of order unity and $h$ not too large \cite{PMHS}. We know that too large an $h$ leads to a different
confining phase with a condensation of spinons and the system moves away from the spin-ice subspace altogether. 
Also, if the disorder $w$ becomes much larger than one, one would once again enter a local phase where the
spins simply point along the local random fields and will no longer be in the spin-ice subspace. The
comparison of our results with the ED study of the random-field model suggests that such a phase is
possible in the RTFIM. The main difference must be that the Ising ordered backbone may be disordered due to the 
local fields. 

It is interesting to consider this study from the perspective of rare-earth magnetic pyrochlore materials,
where RTFIM has been argued to be relevant \cite{savary1,benton,broholm}.
First of all, our work implies that if $h=0$ and one only has weak random fields, the system will
be in a frozen spin-glass phase. A variety of measurements such as NMR or $\mu$SR can easily confirm that.
Indeed restricted subspace generally promotes spin-glass order \cite{chalker}.
This is clearly not the most interesting phase from the point of view of quantum spin-liquids.
However, if disorder becomes a fraction of $J$, then one is away from the perturbative regime and
the possibility of random resonating-hexagon phase becomes likely. While not a true QSL, such a
phase has a lot of quantum fluctuations and local entanglement and should show interesting 
power-law temperature, magnetic
field and frequency dependence \cite{vbg,kimchi} in various responses. These can be dominated by
just the behavior of single hexagons, and can be easily calculated. Indeed this physics can survive
up to very large randomness.

For the material Pr$2$Zr$_2$O$_7$, it has been argued \cite{benton,broholm} that the width of the
random-fields can be much larger than the Ising couplings. The actual distribution of random-fields
was found to not be Gaussian but Lorentzian. But, that should not change the results in a significant way. It
would be interesting if evidence for local resonating hexagons can be seen in these systems.

In order to obtain a true U(1) QSL phase, it is not enough to just reduce the strength of the disorder
or $w$. It is important to have a uniform component of the transverse-field that 
is at least of order the randomness.
What kind of impurities or imperfections can lead to this, if without impurities one has
two-fold degeneracy in the crystal-field states, deserves attention from a materials
point of view. It is more likely that when
non Kramers ions have two nearby non-degenerate crystal-field states
that it can then be modeled as two level systems in a uniform transverse-field. If such a system still
has spin-ice physics, it would be a good candidate for a QSL. Another possibility is that disorder
is correlated over lengths much larger than the lattice constant. In that case, the system
can behave effectively as having a uniform field in any region. Another possibility is that the
magnitude of the random-field is nearly uniform from site to site. But, there is variation in
signs or directions. This model may still have a QSL phase and deserves further attention.

\begin{figure}
\begin{center}
\includegraphics[width=6cm]{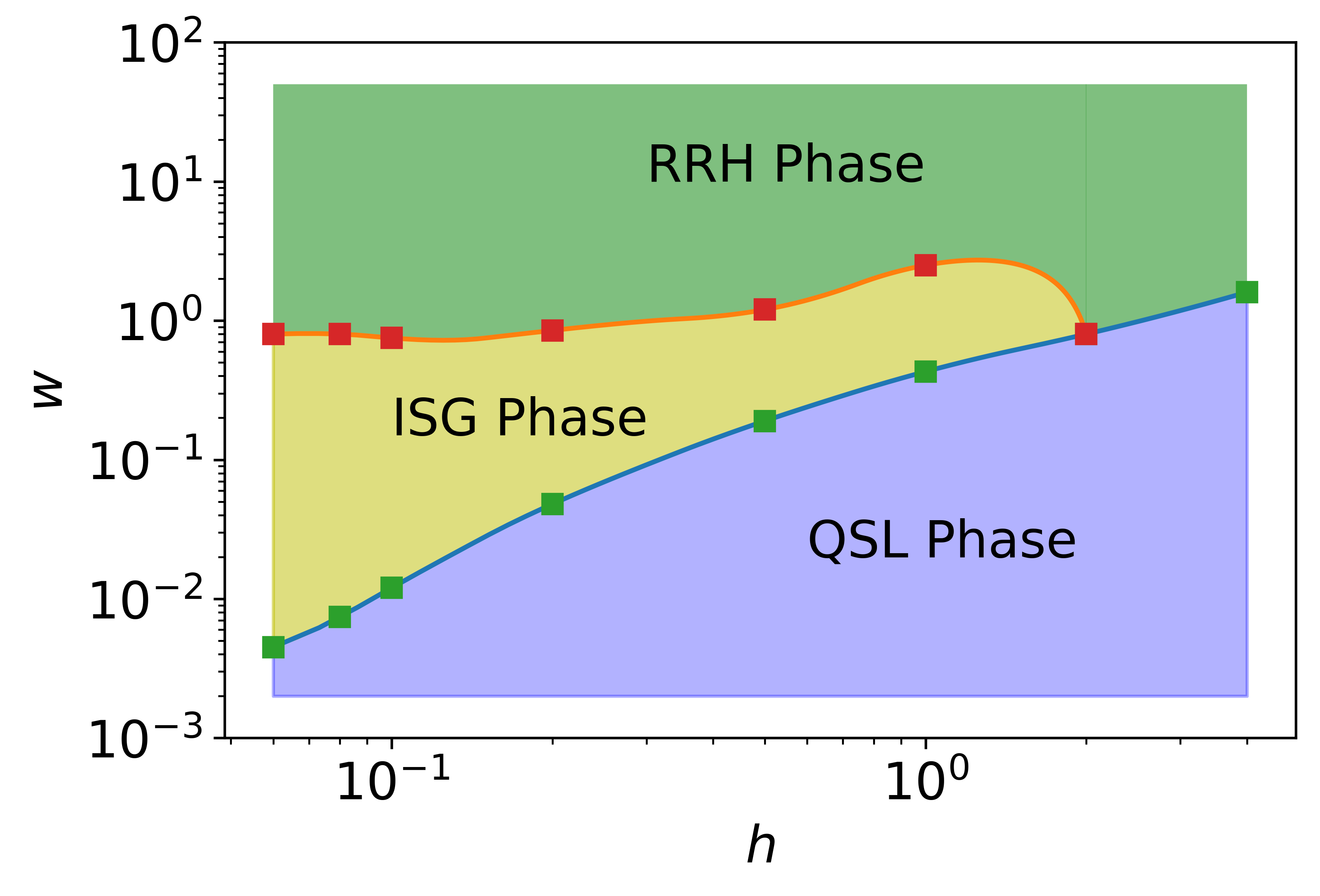}
\caption{\label{Phase} 
	Phase diagram of the model in the $w-h$ plane consisting of quantum spin-liquid (QSL),
	Ising spin-glass (ISG) and random resonating hexagon (RHH) phases.
	}
\end{center}
\end{figure}

\section{Summary and Conclusions}

In summary, in this paper we have studied an effective model derived perturbatively from the random
transverse-field Ising model (RTFIM) on the pyrochlore lattice. The reduced Hilbert space of the effective
model allows us to study larger system sizes and thus deduce the nature of different phases. We find
three different phases. A U(1) QSL phase occurs for sufficiently small randomness. At weak
transverse-fields, increased randomness leads to an Ising spin-glass (ISG) phase, with nearly frozen spins
and very little quantum fluctuations. Increased random-fields can lead to a random resonating hexagon (RRH)
phase, which is a kind of a cluster-glass phase where quantum fluctuations and entanglement are
restricted to small clusters.

We have discussed possible relevance of this study to rare-earth pyrochlores where RTFIM have been argued
to be relevant. It is clear that a broad distribution of transverse-fields, with width exceeding the mean,
will not lead to a QSL phase. But, it is possible for it to still be in a random resonating hexagon phase.
It would be interesting if evidence for local resonating hexagons is observed in Pr$_2$Zr$_2$O$_7$. A true
U(1) QSL would need a material where the magnitude of the transverse-field is nearly uniform at least at nearby
sites.

%\section{Acknowledgements}
Acknowledgements: 
We thank Nic Shannon, Karlo Penc and Stefan Hau-Riege for many helpful discussions.
This work was performed under the auspices of the U.S. Department of Energy by Lawrence Livermore National Laboratory under Contract DE-AC52- 07NA27344.
%This work was performed under the auspices of the U.S. Department of Energy by Lawrence Livermore National Laboratory under Contract DE-AC52-07NA27344. Document release number LLNL-JRNL-778510. 
This work is supported in part by National Science Foundation grant number NSF-DMR 1855111 and 
by grant number NSF-PHY 1748958.

\end{document}